\documentclass[a4paper,9pt,twocolumn]{article}
\usepackage{amsmath}
\usepackage{amsfonts}
\usepackage{slashed}
\usepackage{tensind}
\usepackage{graphicx}
\usepackage{units}

\newcommand{\ii}{\mathrm{i}}

\newcommand{\lnc}{{\Lambda_\text{NC}}}
\newcommand{\hx}{{\hat{x}}}
\newcommand{\AAA}{{\mathcal{A}}}
\newcommand{\hAAA}{{\hat{\mathcal{A}}}}
\newcommand{\komm}[2]{\left[#1,#2\right]}
\newcommand{\moyal}{
	e^{\frac{\ii}{2}\lambda\overset{\leftharpoonup}{\partial}\theta
	\overset{\rightharpoonup}{\partial}}}
\newcommand{\akomm}[2]{\left\{#1 ,#2\right\}}
\newcommand{\nckomm}[2]{\left[#1 \:\overset{\star }{,}\: #2 \right]}

\newcommand{\LL}{\mathcal{L}}
\newcommand{\MM}{\mathcal{M}}
\newcommand{\figref}[1]{Fig.\ref{#1}}
\newcommand{\plotwidth}{5.5cm}
\newcommand{\sun}[1]{\mathbf{SU}\left(#1\right)}
\newcommand{\un}[1]{\mathbf{U}\left(#1\right)}

\title{%
  The Noncommutative Standard Model and Polarization in\\
  Charged Gauge Boson Production at the LHC}

\author{%
  Thorsten Ohl\thanks{e-mail: \texttt{ohl@physik.uni-wuerzburg.de}}$\;{}^{,a}$\\
  Christian Speckner\thanks{e-mail: \texttt{Christian.Speckner@physik.uni-freiburg.de}}$\;{}^{,a,b}$\\
  \hfil\\
  ${}^a$Institut f\"ur Theoretische Physik und Astrophysik\\
  Universit\"at W\"urzburg\\
  Am Hubland, 97074 W\"urzburg, Germany\\
  \hfil\\
  ${}^b$Physikalisches Institut\\
  Albert-Ludwigs-Universit\"at Freiburg\\
  Hermann-Herder-Str.~3, 79104 Freiburg, Germany}

\date{\today}
\begin{document}
\maketitle
\begin{abstract}
We study the pair production of charged gauge bosons at the LHC in a
noncommutative extension of the standard model.  We use angular
distributions in the decays of the gauge bosons to partially
reconstruct polarized cross sections.  We use this, together with $CP$
considerations, to construct more sensitive observables that allow to
separate space-time from space-space noncommutativities.
\end{abstract}

%%%%%%%%%%%%%%%%%%%%%%%%%%%%%%%%%%%%%%%%%%%%%%%%%%%%%%%%%%%%%%%%%%%%%%%%
\section{Introduction}

Following the observation that Yang-Mills theories
on a noncommutative~(NC) spacetime represent a low energy limit of certain open string
theories~\cite{Seiberg:1999vs}, NC gauge theories have
attracted again a lot of
interest. Using Seiberg-Witten Maps~(SWM)~\cite{Seiberg:1999vs} to map such NC gauge
theories\footnote{We will refer to gauge theories living on the NC
manifold as ``NC'' gauge theories (in contrast to ``commutative'' gauge
theories defined on ordinary spacetime), whereas the group structure will be denoted
as ``abelian'' resp.~``nonabelian''.} on Yang-Mills theories living on ordinary spacetime
perturbed by an infinite number of higher dimension operators in a gauge invariant way, a consistent
NC extension of the standard model~(SM) to the Moyal plane has been
constructed~\cite{Madore:2000en,Calmet:2001na,Melic:2005fm}. The Moyal plane
explicitly breaks Lorentz invariance with two constant background
fields $\vec{E}$ and $\vec{B}$ and therefore predicts dramatic new experimental signatures like
the azimuthal dependence of cross sections.

A phenomenological study has been conducted for $Z\gamma$ pair production at the Tevatron and
the LHC~\cite{Alboteanu:2006hh} which suggests that the effects of a NC structure of
spacetime should become visible at the LHC for a new physics scale $\lnc$ up to $\unit[1]{TeV}$.
Since the best lower bounds on $\lnc$ from collider experiments are of the order of $\unit[200]{GeV}$,
this leaves room for a  discovery of noncommutativity at the LHC. The
lower bound on $\lnc$ could possibly be pushed to significantly higher scales by astrophysical
tests as well as precision experiments from atomic physics; however, those bounds assume
noncommutativity on large scales and are model dependent (for other
phenomenological studies of NC
extensions of the SM see~\cite{Hinchliffe:2002km,Behr:2002wx,Ohl:2004tn}). 

In this paper, we study the production of polarized $W$ pairs
at the LHC in the scenario discussed in~\cite{Calmet:2001na}
and present data generated with the Monte-Carlo eventgenerator
WHIZARD~\cite{Kilian:2007gr} for this process
including the semileptonic decay of the $W$ pair. An interesting
feature of the production of $W$ bosons is the possibility of reconstructing
information about their helicity from the angular distribution of the fermionic
decay products. We use this method to reconstruct the
helicity distribution of the $W$ bosons and to construct observables which in principle allow for the
independent measurement of the transverse components of $\vec{E}$ and $\vec{B}$.

The paper is organized as follows: in the second section we briefly review
the construction of the Noncommutative Standard
Model (NCSM) via the Moyal-Weyl product and the application of
Seiberg-Witten maps.
In the third section, we discuss the parton level cross section, with
a particular emphasis on the $CP$ properties that allow the
construction of sensitive observables. In the
fourth section we present results from Monte-Carlo simulations for the production of polarized
$W$ bosons in $pp$ collisions at the LHC and construct
convenient observables. In the fifth section we include the decay of
the $W$ bosons into the simulation and discuss the reconstruction of the
polarized distributions from the semileptonic final state.  In the
appendix we correct an error in the published Feynman rules for the
$f\bar{f}W^+W^-$ vertices in the
NCSM~\cite{Melic:2005fm} and verify our result by the Ward
Identities for the~$q\bar q W^+W^-$ amplitude in the symmetric phase.

\section{The Noncommutative Standard Model}

To construct a NC analog of Minkowski spacetime one considers four noncommuting objects
$\hx^\mu$ with indices raised and lowered using the Minkowski metric $g^{\mu\nu}$ which
transform under Lorentz transformations as a four-vector. The
algebra $\hAAA$ of formal power series in the $\hx^\mu$ is then taken to be the
NC equivalent of the algebra $\AAA$ of functions (fields) on ordinary space time. It is
convenient to parameterize the commutator as
\begin{equation}\label{equ-kommrel}\komm{\hx^\mu}{\hx^\nu}=\frac{\ii}{\lnc^2}\theta^{\mu\nu}\end{equation}
with the new physics scale~$\lnc$.  Below, we will often use the abbreviation $\lambda=\frac{1}{\lnc^2}$.

Due to its antisymmetry and transformation properties, the matrix
$\theta^{\mu\nu}$ can be parameterized with two three-vectors
$\vec{E}$ and $\vec{B}$
\begin{equation*}
\theta^{\mu\nu} = \left(\begin{array}{cccc} 0 & E_x & E_y & E_z 
\\ -E_x & 0 & -B_z & B_y \\ -E_y & B_z & 0 & -B_x \\ -E_z & -B_y & B_x & 0
\end{array}\right)\end{equation*}
which we assume to be constant and commuting in order to obain a Moyal
plane. This implementation
of noncommutativity explicitly breaks Lorentz invariance by introducing two preferred directions in
space.

Instead of working directly with the elements of the algebra $\hAAA$, it is
convenient to represent it in the algebra $\AAA$ of ordinary fields on spacetime with the multiplication
replaced by a deformed product. For the case considered here, this product
is given by the Moyal-Weyl $\star$-product which can be written as
\begin{equation}\label{equ-moyal}\hat{f}(x)\star\hat{g}(x)=\hat{f}(x)\moyal\hat{g}(x) \end{equation}
with the arrow above a partial derivative denots whether it is to
act on $f$ or $g$.
The ``hat'' denotes functions that are to be multiplied with respect to the
Moyal product.  We use the notation $a_\mu b_\nu\theta^{\mu\nu}=a\theta b$ and
$a_\mu\theta^{\mu\nu}=a\theta^\nu$ for contractions of four-vectors with the tensor
$\theta^{\mu\nu}$.

The noncommutativity leads to complications in the construction of Yang-Mills
theories, because the commutator of spacetime dependent gauge
transformation does not close in the Lie algebra. This could be
remedied by extending the Lie algebra to the universal enveloping algebra which also contains the
anticommutators of its elements at the price of adding additional
gauge degrees of freedom.

A very elegant way~\cite{Madore:2000en} to overcome this problem is
provided by Seiberg-Witten maps~\cite{Seiberg:1999vs}. SWMs are maps
from the Lie algebra into the enveloping algebra
\begin{equation*} \tau\longrightarrow\hat{\tau}(\tau,A) \quad,\quad
A^\mu\longrightarrow\hat{A}^\mu(A) \end{equation*}
for the gauge parameter $\tau=\tau_iT_i$ as well as for the gauge
field $A^\mu=A^\mu_iT_i$ such that
commutative gauge transformations
\begin{equation*}A^\mu\xrightarrow{\;\tau\;} A^\mu_\tau \end{equation*}
induce NC gauge transformations
\begin{equation}\label{equ-gauge-equ}\hat{A}^\mu(A)\xrightarrow{\;\tau\;} \hat{A}^\mu(A_\tau)=
\hat{A}^\mu(A)_{\hat{\tau}(\tau,A)} \end{equation}
The gauge equivalence condition \eqref{equ-gauge-equ} can be written in infinitesimal form and, after
expanding the moyal product in the noncommutativity parameter $\lambda$, the
SWMs can be solved for order by order in $\lambda$~\cite{Madore:2000en}.
The solution can be shown to be not unique~\cite{Asakawa:1999cu}; we use the same solution
as~\cite{Calmet:2001na} which preserves the hermicity of the gauge field.
These SWMs are
\begin{gather*}
\hat{\tau}(\tau,A)=\tau + \frac{\lambda}{4}\theta^{\mu\nu}\akomm{\partial_\mu\tau}{A_\nu} +
\mathcal{O}(\lambda^2)\\
\hat{A}^\mu(A)=A^\mu -\frac{\lambda\theta^{\rho\sigma}}{4}\akomm{A_\rho}{\partial_\sigma A^\mu +
{F_\sigma}^\mu} + \mathcal{O}(\lambda^2)\end{gather*}
A similar SWM exists for the matter field $\hat{\Psi}$ and is given by
\begin{equation*} \hat{\Psi}(\Psi,A) = \Psi +
\frac{\lambda\theta^{\mu\nu}}{4}\left(iA_\mu A_\nu\Psi - 2A_\mu\partial_\nu\Psi\right) +
\mathcal{O}(\lambda^2) \end{equation*}

Inserting the SWMs into the NC Yang-Mills action
\begin{multline} S =\int d^4x\:\hat{\LL}_\text{kl} =\int d^4x\:\left(\ii\Hat{\Bar{\Psi}}\star\hat{\slashed{D}}
\star\hat{\Psi} - \right.\\\left.m\Hat{\Bar{\Psi}}\star\hat{\Psi} -
\frac{1}{2}\mathbf{tr}\hat{F}^{\mu\nu}\star\hat{F}_{\mu\nu}\right) \label{equ-lagr-ym}\end{multline}
(which is the commutative Yang-Mills action with all products replaced by
$\star$-products) and expanding the $\star$-product gives a well-defined
effective field theory which incorporates the
noncommutativity of spacetime up to a given order in $\lambda$.
It is gauge invariant order by order in $\lambda$ by virtue of the gauge
equivalence condition \eqref{equ-gauge-equ} and at the same time
contains no new fields in addition to those already present in the commutative case. The new
operators coming from the $\lambda$-expansion are suppressed by powers of $\lnc$.
For this work, only the operators coupling two fermions to one or two
gauge bosons and the operator coupling three gauge bosons are important; the
corresponding Feynman rules for a generic NC Yang-Mills
theory are given in $\lambda$ in the appendix to first order. 

This construction has been used to create a NC extension of the
SM~\cite{Calmet:2001na,Melic:2005fm}. The setup of the fermion and
Higgs sectors
is straightforwardly accomplished by inserting the SWM of the fields into the SM
lagrangian with all products replaced by $\star$-products. The
gauge sector comes with an ambiguity because the coupling of the gauge field
components may be proportional to traces over arbitrary elements of
the enveloping algebra.  In contrast to the commutative case, these
traces are not fixed by the usual normalization of traces over Lie algebra bilinears,
and therefore those couplings depend on the choice of representation matrices. We will work with the
so-called ``non-minimal'' extension of the SM~\cite{Calmet:2001na,Melic:2005fm} 
(\mbox{nmNCSM}). The representation of the SM $\sun{3}\times\sun{2}_\text{L}\times\un{1}$ gauge
group is chosen as the direct sum of the representations belonging to the
fermions and the Higgs boson. The gauge field lagrangian is then written as
\begin{equation*} \LL_\text{GF} = \mathbf{tr} G\hat{F}^{\mu\nu}\hat{F}_{\mu\nu} \end{equation*}
with the field strength given by the covariant derivative of the full product
gauge group (including the gauge couplings)
\[ \hat{F}^{\mu\nu} = \nckomm{\hat{D}^\mu}{\hat{D}^\nu} \]
and a diagonal matrix $G$ which is a Casimir operator of the product group.
$G$ is parameterized by six constants $\frac{1}{g_1^2},\dots,\frac{1}{g_6^2}$
and enforces the correct trace over bilinears of group generators. This condition leaves three free
parameters which are further constrained to the intersection of
simplices~\cite{Behr:2002wx} by the condition $\frac{1}{g_i^2}\ge0$ which ensures
of the positivity of the hamiltonian~\cite{Aschieri:2002mc}. Only one of those
three parameters enters the vertices relevant for this work; in
accordance with~\cite{Behr:2002wx}, we parameterize it as
\begin{equation*} \kappa_2 = \frac{1}{4}\left(\frac{1}{g_5^2}+\frac{1}{g_6^2}-\frac{1}{g_2^2}\right)
\end{equation*}
Plugging the SM couplings into the consistency equations for the $g_i^2$ and putting no
further constraints on them, the simplex condition for $\kappa_2$ reduces to
\begin{equation} -\frac{1}{4g^2}\le\kappa_2\le\frac{1}{4g^2} \label{equ-kappa2}\end{equation}
where $g$ is the isospin gauge coupling.

We will neglect all corrections to the couplings between massive fields and
gauge fields induced by the SWM for the Higgs field. All other Feynman rules
relevant for this work are collected to first order in $\lambda$ in the appendix.

\section{Parton Level Analysis}
\label{sec:parton-level-analysis}

In this section we analyze in detail the partonic process
$\bar{d}d\longrightarrow W^+W^-$ in the NCSM, while the identical discussion for
$\bar{u}u\longrightarrow W^+W^-$ does not need to be repeated.
The scattering amplitude is expanded in $\lambda$
\begin{equation*} \MM=\MM_0 + \lambda\MM_1 + \mathcal{O}(\lambda^2) \end{equation*}
and we study the effects of the NC corrections only to first order in $\lambda$.

For calculating the squared matrix element $\left|\MM\right|^2$, higher orders
should also be truncated to keep the expansion consistent, leading to
\begin{equation}\left|\MM\right|^2 = \left|\MM_0\right|^2 + 2\lambda\Re\MM_0\MM_1^* +
\mathcal{O}(\lambda^2) \label{equ-csect-exp}\end{equation}
In principle, this truncation might yield negative cross sections in some regions
of phase space. There the NC contribution dominates
over the SM amplitude and the inclusion of higher orders in
$\lambda$ is necessary to keep the expansion positive~\cite{Alboteanu:2007bp}.
However, while being a more serious problem at small
$\Lambda_\text{NC}$ and/or high $q\bar{q}$ invariant mass
$\sqrt{s}$~\cite{Alboteanu:2006hh,Alboteanu:2007bp}, this contribution
turns out to be very small and can be safely ignored for the range of
$\Lambda_\text{NC}$ and $\sqrt{s}$ discussed in this work.

At order $\mathcal{O}\left(\lambda^0\right)$, there are two $s$ channel diagrams
and one $t$ channel diagram contributing to this process. At 
$\mathcal{O}\left(\lambda^1\right)$, each of these diagrams receives two
corrections that are obtained by replacing one of the vertices by its
$\mathcal{O}\left(\lambda^1\right)$ correction\footnote{We denote
SM vertices with filled dots and the NC corrections with empty squares.}:\\[3mm]
\centerline{\begin{tabular}{ccc}
\includegraphics{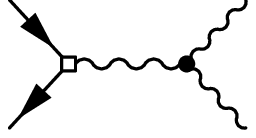} & \qquad &
\includegraphics{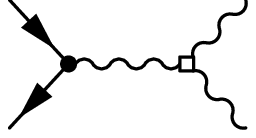} \\
\includegraphics{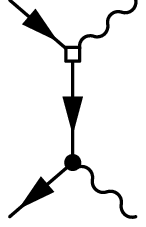} & &
\includegraphics{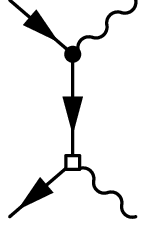}
\end{tabular}}\mbox{}\\[2mm]
In addition, there is a new contact-type diagram which is not present in the SM:
\begin{equation*}\includegraphics{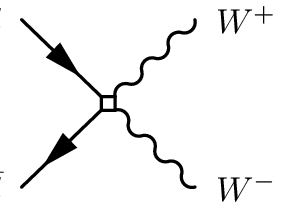}\end{equation*}

In the absence of electroweak symmetry breaking, gauge symmetry forces the
Feynman amplitude to satisfy the Ward identity at all orders in $\lambda$,
specifically at first order, e.g.:
\begin{equation*}\MM_{1_{\scriptstyle\mu\nu}}(k_1,k_2)\epsilon_1^\mu(k_1)k_2^\nu = 0 \end{equation*}
In appendix~\ref{app:ward-dd}, we have checked that this explicitly.
Notably, the $\mathcal{O}(\lambda^1)$ corrections coming
from the vertices coupling three gauge bosons vanish independently of the
other diagrams. This is due to these vertices depending on the group representation
chosen for the gauge sector, which does not influence the other vertices. Also, 
the $s$- and $t$-channel contributions don't cancel completely; the new contact
diagram is necessary for gauge invariance.

In this section, we analyze the process in the center of mass system
(CMS). We label the three-momentum of the $d$ as
$\vec{p}$ and that of the $W^-$ as $\vec{k}$, with $\vec{p}$
oriented in positive $x_3$-direction.

As the introduction of the tensor $\theta^{\mu\nu}$ explicitly breaks
Lorentz invariance, a dependence of the cross section on the azimuthal angle
can be expected to arise. If the external particles are in
helicity eigenstates, the cross section is symmetric under simultaneous
rotations of $\vec{k}$ \emph{and} the NC vectors $\vec{E}$ and $\vec{B}$ around
the $x_3$-axis; therefore, rotating $\vec{k}$ by $\phi$ has the same effect as
rotating the NC vectors by $-\phi$.
Because the expansion of the cross section is truncated after the
first order in $\lambda$ and only contains terms at most linear in
the $E_i$ and $B_i$, the dependence on the azimuthal angle $\phi$ must therefore
be a harmonic oscillation with period $2\pi$. By virtue of the same rotational
covariance, corrections proportional to $E_3$ and $B_3$ must  be independent of
$\phi$.

In the SM, CP symmetry forces the quark spin summed CMS
cross section to be invariant under simultaneous interchange and reversal of
the $W$ helicities
\begin{equation} (h_+,h_-)\xrightarrow{\quad\text{CP}\quad}(-h_-,-h_+)\label{equ-cp-hel}\end{equation}
with~$h_\pm\in\{-1,0,+1\}$ denoting the $W^\pm$ helicities.
The P transformation properties of the commutative gauge theory
carry over to the NC counterpart if $\vec{E}$ transforms as a
vector and $\vec{B}$ as a pseudovector.
However, for the C transformation properties to carry over, an appropriate
choice of SWMs (the maps used in this work satisfy this) and a transformation
of $\theta^{\mu\nu}$ is necessary~\cite{Aschieri:2002mc,SheikhJabbari:2000vi}
\begin{equation*} \theta^{\mu\nu}\xrightarrow{\quad\text{C}\quad}-\theta^{\mu\nu} \end{equation*}
This CP covariance implies that
the NCSM CMS cross section is invariant under the replacement
\begin{equation}
\left.\begin{array}{l} (h_+,h_-) \\ \vec{E} \\ \vec{B}\end{array}\right\}
\xrightarrow{\quad\text{CP}\quad}
\left\{\begin{array}{l} (-h_-,-h_+) \\ \vec{E} \\ -\vec{B}\end{array}\right.
\label{equ-cp-cov}\end{equation}
or, adopting a different point of view, that the CMS cross sections linked by \eqref{equ-cp-hel}
must differ by the sign of $\vec{B}$.
Since cross sections to first order in $\lambda$ can depend at most
linearly on $\vec{B}$, the dependence on~$\vec{B}$ must cancel in that
order after summing over the helicities.

We have calculated the analytical expression for the squared matrix element
summed over helicities and spins using
FORM~\cite{Vermaseren:2000nd} (with the simplification of vanishing $Z$ width).
The result can be put into the form
\begin{equation} \left|\MM\right|^2 = M_0 + \lambda\vec{E}\left(\vec{k}\times\vec{p}\right)M_1
+\mathcal{O}(\lambda^2)\label{equ-analamp} \end{equation}
where $M_0$ and $M_1$ are independent of $\phi$. As argued above, \eqref{equ-analamp} is independent of
$\vec{B}$. In the chosen frame of reference, this can also be written as
\begin{multline}\left|\MM\right|^2 = M_0 + \frac{1}{\lnc^2}
\frac{\sqrt{s}}{2}\left|\vec{k}\right|\sin\theta\;\cdot\\\left(E_1\sin\phi-E_2\cos\phi\right)M_1
+\mathcal{O}(\lambda^2)\label{equ-analamp-comp}\end{multline}
(with the mandelstam $s$) making explicit the harmonic dependence on $\phi$.
\eqref{equ-analamp-comp} shows that the squared amplitude is also independent
of the longitudinal component of $\vec{E}$. We have checked the gauge invariance of our
result by analytically testing the Ward identities for vanishing $Z$ and $W$ mass.

\begin{figure}[tb]\centerline{\includegraphics[width=\plotwidth,angle=270]{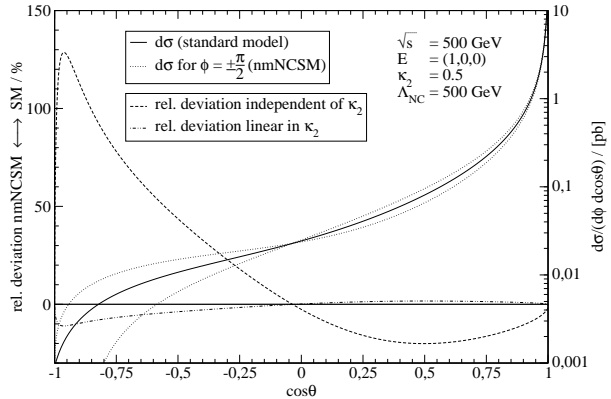}}
\caption{Maximal deviation of $\frac{d\sigma}{d\cos\theta\:d\phi}$ summed over spins and helicities for
the process $d\bar{d}\rightarrow W^+W^-$.}\label{fig-deviation-dd-500}\end{figure}

\figref{fig-deviation-dd-500} shows the NC corrections to the
differential cross section $\frac{d\sigma}{d\cos\theta\:d\phi}$ for $\sqrt{s}=\unit[500]{GeV}$,
$\lnc=\unit[500]{GeV}$ and $\vec{E}=(1,0,0)^T$. Equation \eqref{equ-analamp-comp} shows that for this choice of
$\vec{E}$ the extrema of the azimuthal oscillation are located at $\phi=\pm\frac{\pi}{2}$, so the
plot displays the maximal deviation of the NCSM cross section from the SM prediction. The
deviation has two maxima with a relative deviation of $\approx20\%$ and $\approx120\%$; however, at
these points the differential cross section differs from its maximum by several orders of
magnitude. In the vicinity of the maximum at $\cos\theta=1$ the deviation is only of the order
$<5\%$. Despite $\kappa_2$ being chosen  near the limit of the allowed interval \eqref{equ-kappa2},
the plot shows that the correction proportional to $\kappa_2$ is much smaller than that
independent of $\kappa_2$. This can be understood by noting
that the new parameter only appears in the $s$-channel diagrams,
whereas the cross section is dominated by the $t$-channel diagram for CMS energies high compared with
the $W$ mass. Therefore, we will ignore the effect of $\kappa_2$ for the
remainder of this paper.

We have also implemented the NC vertices and diagrams into a FORTRAN
90 module which we combined with code generated by the matrix element generator
O'Mega~\cite{Moretti:2001zz} to calculate the cross section for the production of
polarized $W$ pairs numerically. Again, we have performed a (numerical) check of
the Ward identities in the limit of vanishing $W$ and $Z$ masses; we have also
checked that the cross section matches the analytical result after helicity
summation.

As a consequence of the finite $Z$ width taken into account in the numerical calculation, the cross
section exhibits a small dependence on the longitudinal component of $\vec{E}$.
Because this correction is smaller than $0.01\%$ of the SM cross section
for reasonable values of the NC parameters, it seems to be of little
experimental relevance.

Turning to the polarized cross sections, we find a dependence on the longitudinal part
of $\vec{B}$ proportional to $\kappa_2$ if the final state consists
a longitudinal and a transverse gauge boson. However, the dominant combination of
helicities for $d\bar{d}\longrightarrow W^+W^-$ is $(+,-)$ (in contrast, for
$u\bar{u}\longrightarrow W^+W^-$ it is $(-,+)$) and all other combinations are suppressed by at
least one order of magnitude for $\sqrt{s}>\unit[300]{GeV}$. Therefore, this effect
is only a small shift in a already suppressed quantity and is unlikely
to be observable in collider experiments. The remaining observable is the azimuthal
oscillation of the cross section.
\begin{figure}[tb]\centerline{\includegraphics[width=\plotwidth,angle=270]{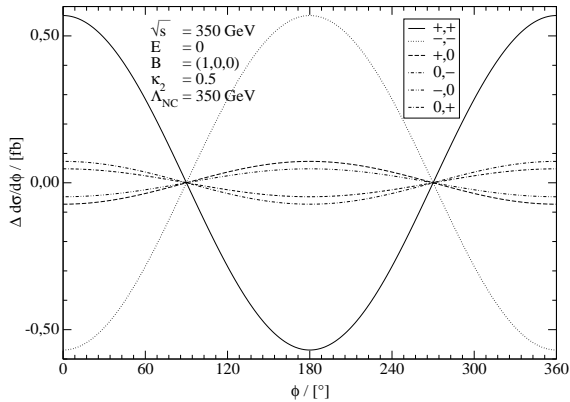}}
\caption{Azimuthal oscillation proportional to $\vec{B}$ of the differential cross section
integrated over $\cos\theta$.}\label{fig-osz-b}\end{figure}

\figref{fig-osz-b} shows the dependence of $d\sigma$ integrated\footnote{Choosing a different
integration interval does not significantly enhance the observed oscillation as $d\sigma$
has a sharp peak at $\cos\theta=1$.} over $\cos\theta$ on $\phi$
for all combinations of helicities for which an azimuthal oscillation proportional to $\vec{B}$ is
allowed at $\mathcal{O}\left(\lambda^1\right)$ by CP covariance~\eqref{equ-cp-cov}. This covariance
is clearly visible and causes the oscillation to
cancel out in the helicity sum. In addition, the oscillation vanishes for final
states containing a longitudinal gauge boson if $\kappa_2$ is set to zero.
Although the oscillation is clearly visible in the plots,
the cross section is strongly suppressed for these
combinations of helicities. Therefore, these channels will be very challenging to
investigate at a collider in the near future.

\begin{figure}[!tb]\centerline{\includegraphics[width=\plotwidth,angle=270]{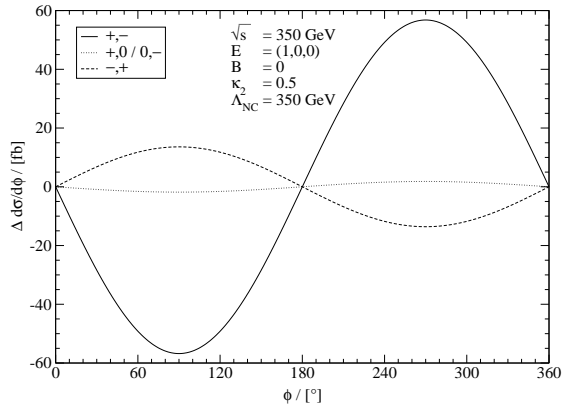}}
\caption{Azimuthal oscillations proportional to $\vec{E}$ of the differential cross section integrated
over $\cos\theta$.}\label{fig-osz-e}\end{figure}
\figref{fig-osz-e} shows the azimuthal oscillation proportional to $\vec{E}$ (combinations for which
the oscillation is barely visible at the chosen scale have been omitted from the plot). Again, the CP
covariance is evident, in this case forcing the cross sections for final states linked by
\eqref{equ-cp-hel} to be identical. In addition, the plot reveals a partial
cancellation in the helicity sum
between the azimuthal oscillation for the combinations $(+,-)$ and $(-,+)$ which reduces the
oscillation exhibited by the summed cross section. Therefore, information about the helicity
distribution of the $W$-bosons should improve observables that probe for NC
extensions of the SM in this process.

\section{Pair Production at the LHC}

We have used the Monte-Carlo eventgenerator WHIZARD~\cite{Kilian:2007gr}
together with custom FORTRAN code to simulate the process $pp\longrightarrow W^+W^-$
at the LHC. A possible problem that haunts this approach is
the truncation of the cross section at order $\mathcal{O}(\lambda^1)$
\eqref{equ-csect-exp} which causes $d\sigma$ to assume
negative values in some (suppressed) regions of phase space as discussed above.
The Monte-Carlo code cannot handle this and therefore the adaption process does
not converge. We have chosen to follow~\cite{Alboteanu:2006hh} and
circumvent this by replacing the
differential cross section $d\sigma$ with
\begin{equation*}\max\left\{d\sigma,0\right\}  \end{equation*}
which is supported by second order calculations~\cite{Alboteanu:2007bp}.
We have used the CTEQ5M~\cite{Lai:1999wy} series of PDFs with a running
factorization scale equal to $\sqrt{s}$. The eventgenerator keeps track
of the helicities of the $W$ bosons; we used this
information to discriminate between the $W$ helicities in the analysis presented
in this section.

We generated $1.5\cdot10^6$ events
(which corresponds to an integrated luminosity of $\int dt\:\LL\approx\unit[20]{\text{fb}^{-1}}$)
with a hadronic CMS energy of $\unit[14]{TeV}$ in each run and applied a cut
\begin{equation} \unit[200]GeV\le\sqrt{s}\le\unit[1]{TeV}\label{equ-cut-s}\end{equation}
that removes events near the production threshold (which are supposedly more contaminated with
background processes) as well as events with very high CMS energy (for
which the cancellation of the expansion at $\mathcal{O}(\lambda^1)$ may not be
justified). Since any acceptance cuts on the polar angle don't affect the $W$-momenta
directly but rather those of the decay products, we didn't apply any such cuts.

As the partonic initial states contain one valence quark (the quark) and
one sea quark (the antiquark), the partonic CMS system generally is strongly
boosted  with respect to the laboratory system. This boost mixes the
transversal components of the NC vectors according to
\begin{align*} E_1&\xrightarrow{\;\Lambda\;}\gamma(E_1-\beta B_2) \quad
&B_1&\xrightarrow{\;\Lambda\;}\gamma(B_1+\beta E_2)\\
E_2&\xrightarrow{\;\Lambda\;}\gamma(E_2+\beta B_1) & B_2&\xrightarrow{\;\Lambda\;}
\gamma(B_2-\beta E_1)\end{align*}
(with the usual Lorentz factors $\beta$ and $\gamma$) and also mixes the polarization vectors. As a
result, $\vec{B}$ in the laboratory frame gets mapped to $\vec{E}$ in the CMS and therefore its
influence on the cross section may be much greater than the CMS result of
the last section suggests.

The initial state contains two identical particles, and therefore
the azimuthal distribution must be invariant under rotations by $\pi$
around the $x_1$ and $x_2$ axes if no asymmetric polar cuts
are made (again simultaneously rotating $\vec{E}$ and $\vec{B}$).
This means invariance under the replacements
\begin{equation*} E_1\rightarrow-E_1 \quad B_1\rightarrow-B_1 \quad \phi\rightarrow\pi-\phi
\end{equation*}
as well as
\begin{equation*} E_2\rightarrow-E_2 \quad B_2\rightarrow-B_2 \quad \phi\rightarrow-\phi
\end{equation*}
implying that any observable azimuthal oscillation caused by $E_1$ and $B_1$ must be proportional to
$\cos\phi$, while parts proportional to $\sin\phi$ must cancel between events with antiquarks
coming from negative $x_3$ direction and those with antiquarks coming from
positive $x_3$ direction. Therefore, as \figref{fig-osz-e} shows that the
oscillation caused by $E_1$ is proportional to $\sin\phi$, one would expect it
to cancel out, while $B_1$ (getting mapped to $E_2$) should cause an oscillation
proportional to $\cos\phi$.

Examining the azimuthal distributions without cuts on the polar angle $\theta$, one does
indeed find such an oscillation proportional to $\vec{B}_\perp$,
but none proportional to $\vec{E}_\perp$. To observe such an effect, one has
to apply appropriate cuts that favor antiparticles coming from either
positive or negative $x_3$ direction~\cite{Alboteanu:2006hh}.
One can easily convince oneself that the cut
\begin{equation} 0\le(\theta_-+\theta_+)\le\pi \label{equ-cut-theta-1}\end{equation}
favors events with antiquark momenta pointing in negative $x_3$ direction, while the
complementary cut
\begin{equation} \pi\le(\theta_-+\theta_+)\le2\pi \label{equ-cut-theta-2}\end{equation}
favors the opposite case.

\begin{figure}[!tb]\centerline{\includegraphics[width=\plotwidth,angle=270]{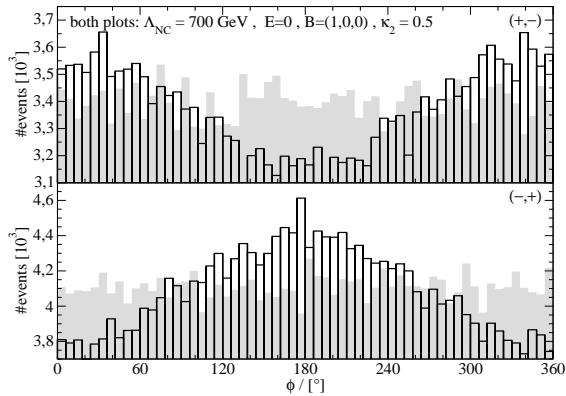}}
\caption{Oscillation of the azimuthal distribution proportional to $\vec{B}_\perp$; no cuts on the
polar angle.}\label{fig-adist-wwb}\end{figure}
\begin{figure}[!bt]\centerline{\includegraphics[width=\plotwidth,angle=270]{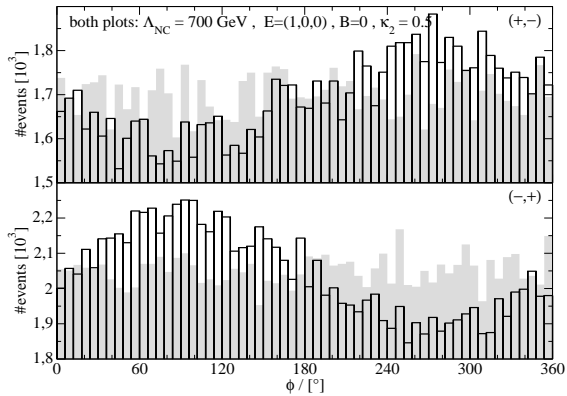}}
\caption{Oscillation of the azimuthal distribution proportional to $\vec{E}_\perp$ after applying
cut \eqref{equ-cut-theta-1}.}\label{fig-adist-wwe}\end{figure}

\figref{fig-adist-wwb} shows the azimuthal distribution
for $\lnc=\unit[700]{GeV}$ compared with the SM prediction for the helicity
combinations $(-,+)$ and $(+,-)$ and $\vec{B}=(1,0,0)^T$, $\vec{E}=0$.
All Monte-Carlo distributions show the
azimuthal angle of the $W^-$ (that of the $W^+$ differs just by a shift of $\pi$). The gray
distributions are the SM expectation.
Both distributions exhibit a significant
oscillation with a relative phase shift of $\pi$; the total event count is larger for
the combination $(-,+)$. The azimuthal distribution for the case of $\vec{B}=0$, $\vec{E}=(1,0,0)^T$
after application of cut \eqref{equ-cut-theta-1} is displayed in \figref{fig-adist-wwe}. Again, a
rather large oscillation can be observed with a phase shift of $\pi$ between $(-,+)$ and
$(+,-)$. This oscillation is \emph{not} visible if no asymmetric cuts on the polar angle are
applied.

There is a similar azimuthal oscillation proportional to $\vec{E}_\perp$ and $\vec{B}_\perp$ for the
combination $(-,-)$; however, the statistics is much lower for this combination
of helicities. Finally, there are no visible oscillations for $(+,+)$ and
the SM cross section is much smaller for this combination anyway.
The simulated
distributions show no visible dependence on the longitudinal components of the NC vectors which is
consistent with the fact of those contributions to the partonic cross section being very small (the
longitudinal components of $\vec{E}$ and $\vec{B}$ are not affected by the boost from the laboratory
frame into the CMS).

Since the azimuthal oscillation proportional to $\vec{E}_\perp$ cancels out if no cut on $\theta$ is
applied, the oscillations for the two cuts \eqref{equ-cut-theta-1} and \eqref{equ-cut-theta-2} must
differ only by a phase shift of $\pi$ causing them to cancel out in the sum. Therefore, taking those
two distributions, shifting one of them by $\pi$ and then summing them up will
cause the oscillations
proportional to $\vec{E}_\perp$ to add up constructively and increase the statistics by a factor of
two. In addition, the simulations show that the oscillation proportional to $\vec{B}_\perp$ is the
same for the cuts \eqref{equ-cut-theta-1} and \eqref{equ-cut-theta-2} and therefore
cancels in this sum. This way, at first order in $\lambda$, measuring the
azimuthal distribution without any (or with symmetric) cuts on $\theta$ in principle allows for the
determination of $\vec{B}_\perp$, while measuring the azimuthal distributions for the cuts
\eqref{equ-cut-theta-1}, \eqref{equ-cut-theta-2} and adding them with a shift of $\pi$ filters
out the oscillation proportional to $\vec{B}_\perp$ and allows for an independent measurement of $\vec{E}_\perp$.

Note that our results differ from those presented
in~\cite{Conley:2008kn}, which predicts a stronger dependence
on~$\vec{E}_\perp$ effect at the LHC.  We can reproduce
the numbers of~\cite{Conley:2008kn} using the $f\bar{f}W^+W^-$
Feynman rule~(85) in~\cite{Melic:2005fm} both for $u\bar{u}W^+W^-$ and
for $d\bar{d}W^+W^-$ vertices.  However, 
the correct vertices in appendix~\ref{app:nmNCSM-Feynman-rules} have different
momentum dependences.  We prove this in appendix~\ref{app:ward-dd} by
checking the Ward Identities for~$q\bar qW^+W^-$ in the symmetric phase.

\section{Semileptonic Decay and Helicity Reconstruction}

In order to apply the observables constructed in the last section
to real measurements at the LHC, we must account for the fact that 
$W$ bosons are unstable and decay.  Specifically, we will concentrate on the
semileptonic decay channel $pp\rightarrow W^+W^-\rightarrow e\bar{\nu}_eu\bar{d}$.

In the narrow width approximation, integrating over the azimuthal angles of
the fermionic decay products (with the momentum of the parent $W$ boson as polar
axis), the cross section decomposes as
\begin{multline*} \frac{d\sigma}{d^3k_+\:d^3k_-\:d\bar{\theta}_+\:d\bar{\theta}_-} 
\propto \\\sum_{h_+h_-}\frac{d\sigma_{h_+h_-}}{d^3k_+\:d^3k_-}P_{h_+}\left(\cos\bar{\theta}_+\right)
P_{h_-}\left(\cos\bar{\theta}_-\right) \end{multline*}
where the $k_\pm$ are the four-momenta of the intermediary $W$ bosons, $\bar{\theta}_\pm$ are the polar
angles of the $\bar{\nu}_e$ resp. $\bar{d}$ in the CMS of the $W$s,
$d\sigma_{h_+h_-}$ is the cross section for the production of polarized $W$ bosons and the $P_h$ are polynomials
\begin{equation*}
P_h(x)=\begin{cases}\frac{1}{2}(1\pm x)^2 & \text{for $h=\pm1$} \\
                     1-x^2 & \text{for $h=0$} \end{cases} \end{equation*}
The functions
\begin{equation*}Q_h(x)=\begin{cases} -\frac{1}{2}\pm x+\frac{5}{2}x^2 &  \text{for $h=\pm1$} \\
2-5x^2 & \text{for $h=0$} \end{cases} \end{equation*}
have the projection property
\begin{equation*} \int_{-1}^1dx\:Q_h(x)P_{h'}(x)=\delta_{hh'}\int_{-1}^{1}dx\:P_{h'}(x) \end{equation*}
Therefore, it is possible to obtain the cross section for the production
of polarized $W$ bosons by convoluting the cross section for
$pp\rightarrow W^+W^-\rightarrow e\bar{\nu}_eu\bar{d}$ with the $Q_r$,
projecting out the contribution of a certain helicity combination to the
cross section. On the level of counting events, this can be done by binning the
events with weights
\begin{equation*} w_{h_+h_-}=Q_{h_+}\left(\cos\bar{\theta}_+\right)Q_{h_-}\left(\cos\bar{\theta}_-\right)
\end{equation*}
The resulting distributions of the weighted events then reproduce the
pair production distributions for the helicity combination $(h_+,h_-)$,
and summing over the helicity
indices reproduces the unweighted distributions by virtue of the
normalization of the $Q_h$.

However, measuring the charge of a quark jet is
virtually impossible, and therefore the corresponding angle $\bar{\theta}$ can
only be measured up to a sign. Since
$Q_0(x)=Q_0(-x)$ and $Q_{\pm1}(-x)=Q_{\mp1}(x)$, this boils down to not being able to discriminate between
the transverse polarizations of the hadronically decaying $W$.

If the number of events in a bin $N(k_i^\pm,x^+_i,x^-_j)$ (with the
abbreviation $x^\pm$ for $\cos\bar{\theta}_\pm$) is distributed with a statistical error $\Delta N=\sqrt{N}$,
then the statistical error $\Delta N_{h_+h_-}$ of the corresponding bin $N_{h_+h_-}$ in the
reconstructed distribution is given by the geometric mean
\begin{multline} \Delta N(k^\pm_l)_{h_+h_-} = \\
\sqrt{\sum_{ij} N(k^\pm_l,x_i^+,x_j^-)Q_r(x_i^+)^2 Q_s(x_j^-)^2} \label{equ-rec-error}\end{multline}
where the sum runs over some division of the allowed range of $\cos\bar{\theta}_\pm$ (or over single
events). Explicit calculation shows
\begin{equation*} \int_{-1}^1dx\:P_{h}(x)Q_{h'}(x)^2 > \int_{-1}^1dx\:P_{h}(x)Q_{h'}(x) \end{equation*}
for all $h,h'$ and therefore the statistical error of the reconstructed distribution will
be greater than $\sqrt{N_{h_+h_-}}$.

As the neutrino coming from the leptonically decaying $W$ cannot be detected
directly, its momentum has to be reconstructed from the other momenta and
$p_{T,\text{miss}}$. This can be done using the mass shell conditions of the $W$ and
that of the neutrino; see~\cite{Ohl:2008ri} for more details. However, this
procedure gives two solutions for neutrino momentum, one of which
approximates the correct momentum for events that can be assigned to a decaying
$W$.

There is no simple way to discriminate between those two solutions; the simulated
data shows that one has in fact to choose between them on a per event basis to get
a reasonable reconstruction of the neutrino momentum. We have found that choosing
the solution that minimizes the cosine of the angle between the two $W$ momenta
allows to reconstruct at least $60\%$ of the momenta correctly. By
a ``correctly'' reconstructed momentum we mean the solution with $p^3$
being closest to the ``truth'' neutrino momentum available from Monte-Carlo data.
Also, simulation shows that this method of choosing solutions preserves the
oscillations discussed in the last section and still allows for the
independent measurement of $\vec{E}_\perp$ and $\vec{B}_\perp$ (this is
nontrivial since incorrectly reconstructed events might mix up the complementary
distributions that combine to cancel out the oscillations proportional to
$\vec{E}_\perp$ resp. $\vec{B}_\perp$). Therefore, we have chosen to use this
criterion to disambiguate the two solutions in our analysis\footnote{Another
possibility of dealing with the two solutions would be to count them both into
the histogram like in~\cite{Ohl:2008ri}, obtaining better signal statistics at
the price of more noise.}.

In addition to the cascade diagrams, there are 17 additional SM
diagrams contributing to the matrix element for
$d\bar{d}\longrightarrow e\bar{\nu}_eu\bar{d}$ and similarly for
$u\bar{u}\longrightarrow e\bar{\nu}_eu\bar{d}$. These background diagrams contain
one separate gauge equivalence class~\cite{Boos:1999qc} which consists of all
diagrams with one quark line connecting initial and finals states and which we chose to ignore
as these topologies can be strongly suppressed with invariant mass cuts.
As the other diagrams are necessary for gauge invariance, we have
implemented them together with the cascade type diagrams into an event generator
using WHIZARD.

We have generated events for an integrated luminosity of $\int
dt\:\LL=\unit[400]{\text{fb}^{-1}}$ and have also included the cases of the
quark pair being $c\bar{s}$ and the lepton-neutrino pair being one  of
$\mu^-\bar{\nu}_\mu\;,\;\tau^-\bar{\nu}_\tau$ with a naive factor of 6 on the
integrated luminosity\footnote{Statistics can be improved
by an additional factor $2$ if semileptonic decays of the type $pp\rightarrow
e^+\nu_e \bar{u}d$ are also included.}.

In addition to the cut
\eqref{equ-cut-s} on $\sqrt{s}$ we also applied an acceptance cut
\begin{equation} 5^\circ\le\theta\le 175^\circ \label{equ-cut-theta}\end{equation}
to all particle momenta (with the exception of the neutrino momentum) and a cut
\begin{equation*} \unit[70]{GeV}\le m_+\le \unit[90]{GeV} \end{equation*}
on the invariant mass of the jet (quark) pair.

\begin{figure}[!tb]\centerline{\includegraphics[width=\plotwidth,angle=270]{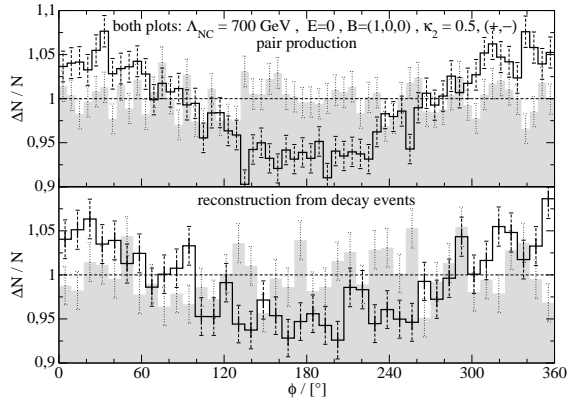}}
\caption{Comparison of the azimuthal distribution of $1.5\cdot 10^6$ simulated
events for $W$ pair production with that of
events reconstructed from the simulation of $pp\longrightarrow e\bar{\nu}_eu\bar{d}$
($\approx3.7\cdot10^6$ events).}\label{fig-errors}\end{figure}
\figref{fig-errors} shows an azimuthal distribution obtained from the simulated production of
polarized $W$ pairs versus the corresponding distribution obtained after reconstructing the
helicity distribution from the semileptonic decay (the neutrino momentum has been directly extracted
from Monte-Carlo data for this distribution). In the first case, the error bars have been calculated
as $\sqrt{N}$, in the second case according to \eqref{equ-rec-error}.
Although the statistics is much better for the second
case ($\approx 3.7\cdot 10^6$ events in contrast to $1.5\cdot 10^6$) and the number of bins has been
reduced from $60$ to $40$, the statistical error is of the same order than that in the first case.
This fact is further emphasized by the thick black error bars in the second distribution which show
the errors to be expected if the events were distributed as $\sqrt{N}$ and which are much smaller than the
actual fluctuations.

As stated above, it is not possible to discriminate between the transverse
polarizations of the hadronically decaying $W^+$, so we sum over both, denoting the
sum of distributions for the helicities $(+,-)$ and $(+,+)$ as $(+,\mp)$
and the sum of $(-,+)$ and $(-,-)$ as $(-,\pm)$. The distributions
\figref{fig-adist-wwb} and \figref{fig-adist-wwe} show that the relative
oscillations for $(+,-)$ and $(-,+)$ differ only by a shift of $\pi$, so we
can enhance the observables further by shifting the azimuthal distribution
for $(-,\pm)$ by a phase of $\pi$ and adding it to that for $(+,\mp)$.

To put together all the steps in our final analysis: we reconstruct the neutrino
momenta and bin the events weighted with the projector functions
to obtain azimuthal distributions
for the production of polarized $W$ bosons. Adding the distributions for
$(+,\mp)$ and $(-,\pm)$ with a phase shift of $\pi$ and without any polar cuts
apart from \eqref{equ-cut-theta}, we obtain an observable which is sensitive
only to $\vec{B}_\perp$ and \emph{not} to $\vec{E}$, the oscillation of which we
will call $\vec{B}$-oscillation. Taking the same sum for each of the two
cuts \eqref{equ-cut-theta-1} and \eqref{equ-cut-theta-2} and adding those
distributions with a phase shift of $\pi$, we obtain another observable which
is sensitive to $\vec{E}_\perp$ only, the oscillation of which we will call
$\vec{E}$-oscillation. The phase of these oscillations can be used to determine
the alignment of $\vec{E}_\perp$ and $\vec{B}_\perp$ in the plane perpendicular
to the beam axis, while their magnitude contains information about the absolute
values.
\begin{figure}[!tb]\centerline{\includegraphics[width=\plotwidth,angle=270]{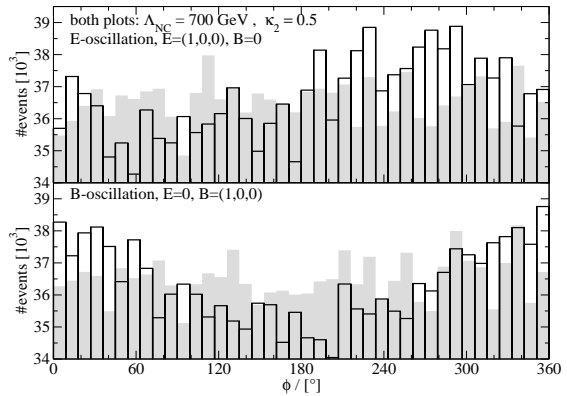}}\caption{Azimuthal
oscillations at the LHC after reconstruction of the momenta and the helicity distributions for
$\lnc=\unit[700]{GeV}$; $\int dt\:\LL=\unit[400]{\text{fb}^{-1}}$.}\label{fig-aosc-700}\end{figure}

\begin{figure}[!tb]\centerline{\includegraphics[width=\plotwidth,angle=270]{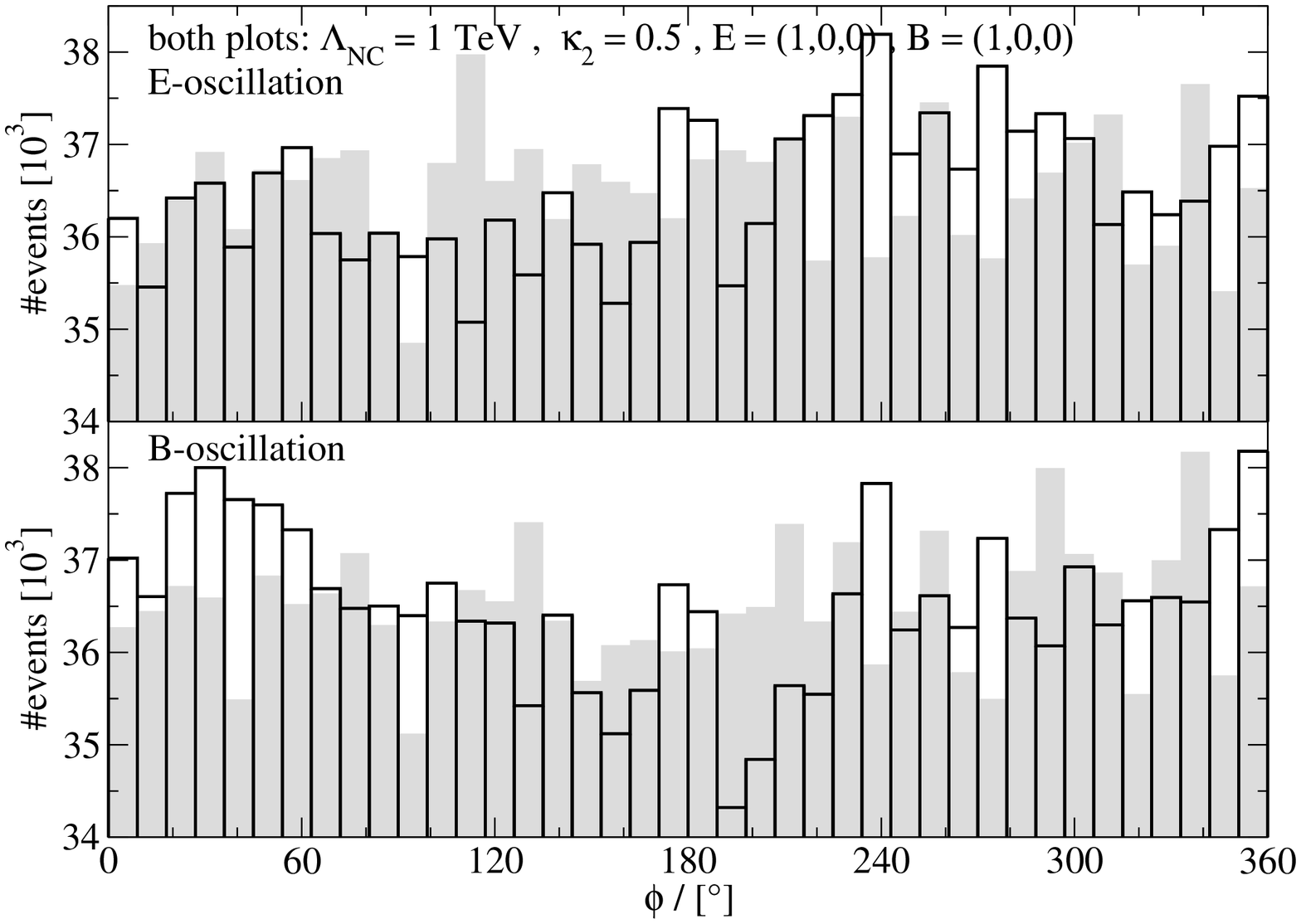}}\caption{Like
\figref{fig-aosc-1000} for $\lnc=\unit[1]{TeV}$.}\label{fig-aosc-1000}\end{figure}

\figref{fig-aosc-700} and \figref{fig-aosc-1000} show these observables for
$\lnc=\unit[700]{GeV}$ and $\lnc=\unit[1]{TeV}$. For $\lnc=\unit[700]{GeV}$, the
azimuthal oscillation is clearly visible, although comparison with
\figref{fig-adist-wwb} and \figref{fig-adist-wwe} shows that the statistical
error of the distribution is much larger than for the hypothetical case
of $W$ pair production (even though we simulated $\approx 3.7\cdot10^6$
events as opposed to the $1.5\cdot 10^6$ events in the pair production
simulation). For $\lnc=\unit[1]{TeV}$, the oscillation is still visible.

To get a quantitative handle on the significance of the oscillations, we have
done a naive calculation of $\chi^2$ with respect to the SM
prediction for the histograms \figref{fig-aosc-700} and \figref{fig-aosc-1000}.
For $\lnc=\unit[700]{GeV}$, this shows a clear deviation from the standard
model. Unfortunately, for $\lnc=\unit[1]{TeV}$, this naive $\chi^2$ test shows no
clear deviation from the SM, suggesting $\unit[1]{TeV}$ might
be the limit of the LHC discovery reach in this channel (consistent with the
results obtained in~\cite{Alboteanu:2006hh}). The analysis might be improved over
the $\chi^2$ test by considering a quantity more sensitive to the harmonic
oscillation exhibited by the distribution, e.g. dividing the histogram into two
hemispheres and looking at the difference in the event count. However, such an
analysis would be intimately tied to the issue of disentangling the movement of
the collider frame of reference with respect to the rest frame of
$\theta^{\mu\nu}$ from the measurement and therefore lies outside of the scope of this work.

\section{Conclusions}
We have studied the pair production of charged gauge bosons at the LHC
in the NCSM.  From the angular distributions in the decays of the
gauge bosons we are able to partially reconstruct polarized cross
sections.  Using $CP$ transformation properties we have constructed
sensitive observable from suitable combinations of distributions.
While the physics reach of the LHC remains at the TeraScale, these
observables allow us to separate space-time from space-space
noncommutativities.

\section*{Acknowledgements}
This research is supported by Bundesministerium f\"ur Bildung und
Forschung Germany, grant 05HT6WWA and by the Helmholtz Alliance
\textit{Physics at the Terascale}.  CS is supported by Deutsche
Forschungsgemeinschaft through the Research Training Groups GRK\,1147
\textit{Theoretical Astrophysics and Particle Physics} and  GRK\,1102
\textit{Physics of Hadron Accelerators}.

\begin{appendix}

\section{Feynman rules}

\subsection{General Yang Mills theory}
\label{app:fr-yang-mills}

Collected in this section are the Feynman rules relevant for this work for a
generic Yang Mills theory. As far as we know, they have not been previously
presented in the literature. Feynman rules containing group generator matrices
are to be understood as taking the matrix element corresponding to the
combination of fermions meeting at the vertex. All momenta are ingoing.
The covariant derivative is chosen as
\begin{equation*} D^\mu\Psi = \partial^\mu\Psi-igA^\mu\Psi \end{equation*}

The expansion of the lagrangian \eqref{equ-lagr-ym} to first order in $\lambda$ yields the following
correction to the vertex $\bar{\Psi}\Psi A$
\begin{equation*}
\parbox{33mm}{\includegraphics{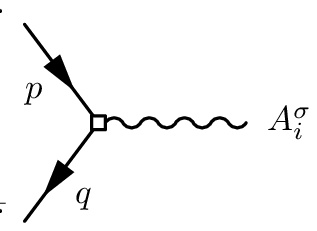}}\quad=\quad
\frac{\lambda}{2}gT_i\theta^{\mu\nu\sigma}p_\mu q_\nu
\end{equation*}
with the totally antisymmetric symbol
\begin{equation*} \theta^{\mu\nu\sigma} = \theta^{\mu\nu}\gamma^\sigma +
\theta^{\nu\sigma}\gamma^\mu + \theta^{\sigma\mu}\gamma^\nu \end{equation*}
The vertex $AAA$ receives the correction
\begin{multline*}
\parbox{35mm}{\includegraphics{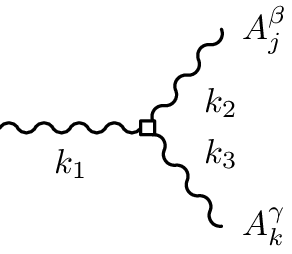}}\quad=\\
-\lambda gT_{i\left\{jk\right\}}\,\theta\left((k_1,\alpha),(k_2,\beta),(k_3,\gamma)\right)
\end{multline*}
with the kinematic factor
\begin{multline*}
\theta\left((k_1,\alpha),(k_2,\beta),(k_3,\gamma)\right)\quad=\\\begin{aligned}
\qquad\quad&\theta_{\alpha\beta}((k_2k_3)k_{1\gamma} - (k_1k_3)k_{2\gamma})+\\&
k_1\theta k_2(k_{3\beta}g_{\alpha\gamma}-k_{3\alpha}g_{\beta\gamma}) {}-{} \\&
k_1\theta_\alpha (k_{2\gamma}k_{3\beta} - (k_2k_3)g_{\beta\gamma}) +\\&
k_1\theta_\beta(k_{2\gamma}k_{3\alpha} -(k_2k_3)g_{\alpha\gamma})+ \\&
k_1\theta_\gamma(k_{3\beta}k_{2\alpha}-(k_3k_2)g_{\alpha\beta}) +\end{aligned}\\
\mbox{cycl. perm. of} \left((\alpha,1),(\beta,2),(\gamma,3)\right)
\end{multline*}
and the representation dependent trace
\begin{equation*} T_{i\{jk\}}=\mathbf{tr} T_i\akomm{T_j}{T_k}=\frac{1}{3}\sum_{\text{perm}}T_iT_jT_k
\end{equation*}
A straightforward calculation shows that the vertex factor vanishes by contraction with one of the
three momenta as required by gauge invariance.

\eqref{equ-lagr-ym} also gives rise to a new vertex $\bar{\Psi}\Psi AA$
\begin{multline*}
\parbox{30mm}{\includegraphics{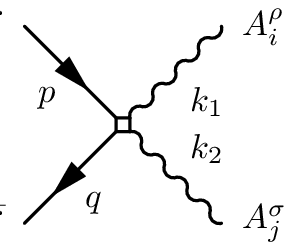}}
\quad=\\
\frac{\lambda}{2}g^2\left(\slashed{p}\theta^{\sigma\rho}\komm{T_i}{T_j} +\right.\\
\left(\slashed{k}_1\theta^{\sigma\rho}+p\theta^\sigma\gamma^\rho +
q\theta^\rho\gamma^\sigma\right)T_i T_j + \\
\left(\slashed{k}_2\theta^{\rho\sigma} + p\theta^\rho\gamma^\sigma
+ q\theta^\sigma\gamma^\rho\right)T_j T_i +\\\left.
\left(k_1\theta^\sigma\gamma^\rho + k_2\theta^\rho\gamma^\sigma
\right)\akomm{T_i}{T_j}\right)
\end{multline*}

\subsection{nmNCSM}
\label{app:nmNCSM-Feynman-rules}
From the general rules present above, the nmNCSM Feynman rules relevant for
this work can be readily calculated by plugging in the nmNCSM fields, group
representations and gauge couplings. While most of the following rules can
already be found in~\cite{Melic:2005fm}, we repeat our result here because
the $f\bar f W^+W^-$ Feynman rule given in~\cite{Melic:2005fm} applies only to
the isospin up case, while the isospin down
case turns out to be different. In
appendix~\ref{app:ward-dd} we demonstrate that this variation is
required by the corresponding Ward identities.

For all other rules, our calculation agrees with the
results published in~\cite{Melic:2005fm}, and our isospin down $f\bar f W^+W^-$ coupling also
agrees with the Lagrangian given there (care must be taken in the comparision as
we give the rules for an ingoing fermion and an ingoing antifermion, while the
authors of~\cite{Melic:2005fm} chose an ingoing and an outgoing fermion).

\begin{equation*}
\parbox{33mm}{\includegraphics{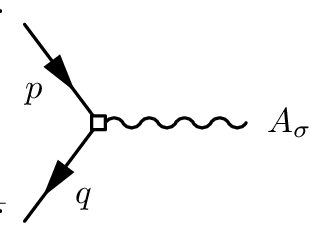}}
\quad=\quad\frac{\lambda e}{2}Qp_\mu q_\nu\theta^{\mu\nu\sigma}\end{equation*}
\begin{multline*}
\parbox{40mm}{\includegraphics{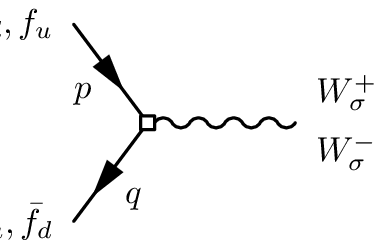}}\quad=\\
\frac{\lambda g}{2\sqrt{2}}p_\mu q_\nu\theta^{\mu\nu\sigma}\frac{1-\gamma^5}{2}\end{multline*}
(where $f_u$ resp. $f_d$ denote a fermion with isospin up resp. down)
\begin{multline*}
\parbox{33mm}{\includegraphics{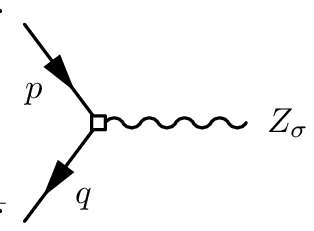}}
\quad=\\\frac{\lambda e}{2\sin(2\theta_W)}p_\mu q_\nu\theta^{\mu\nu\rho}(g^f_V - g^f_A\gamma^5)
\end{multline*}
with the vector and axial couplings
\begin{align*} g^d_V&=\frac{2}{3}\sin^2(\theta_W)-\frac{1}{2}\qquad\qquad & g^d_A&=-\frac{1}{2} \\
g^u_V&=\frac{1}{2}-\frac{4}{3}\sin^2(\theta_W) & g^u_A&=\frac{1}{2} \\
g^l_V&=2\sin^2\theta_W-\frac{1}{2} & g^l_A&=-\frac{1}{2} \\
g^\nu_V&=\frac{1}{2} & g^\nu_A&=\frac{1}{2}
\end{align*}
\begin{multline*}
\parbox{30mm}{\includegraphics{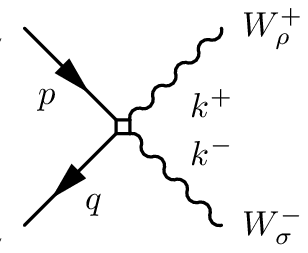}}\quad=\\
\frac{g^2}{4}\theta^{\sigma\rho\mu}(k^++p)_\mu\frac{1-\gamma^5}{2}
\end{multline*}
\begin{multline*}
\parbox{30mm}{\includegraphics{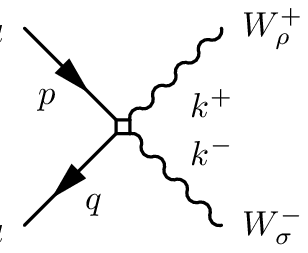}}\quad=\\
\frac{g^2}{4}\theta^{\sigma\rho\mu}(k^++q)_\mu\frac{1-\gamma^5}{2}
\end{multline*}
\begin{multline*}
\parbox{35mm}{\includegraphics{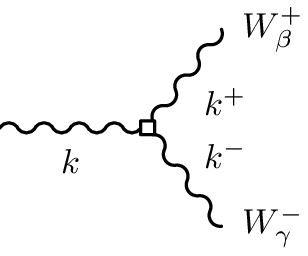}}\quad=\\
-2eg^2\kappa_2\lambda\,\theta\left((k,\alpha),(k^+,\beta),(k^-,\gamma)\right)
\end{multline*}
\begin{multline*}
\parbox{35mm}{\includegraphics{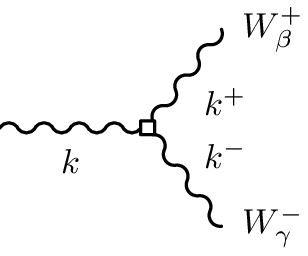}}\quad=\\
2egg'\kappa_2\lambda\,\theta\left((k,\alpha),(k^+,\beta),(k^-,\gamma)\right)
\end{multline*}

\section{Ward identities for $u\bar{u}\longrightarrow W^+ W^-$ and  $d\bar{d}\longrightarrow W^+ W^-$}
\label{app:ward-dd}

We demonstrate the necessity of amending the Feynman rules
of~\cite{Melic:2005fm} as in appendix~\ref{app:nmNCSM-Feynman-rules}
by verifying the corresponding Ward identity
in the limit of vanishing
gauge boson masses:
\begin{equation}
\label{eq:WI}
  k^+_\mu\epsilon^-_\nu\mathcal{M}^{\mu\nu}(q\bar q\to W^+W^-) = 0\,.
\end{equation}
Since the full $\sun{1}_L\times\un{1}_Y$ symmetry is restored in this limit
, we can simplify the calculation by using the gauge
eigenstate~$W_3$ instead of~$A$ and~$Z$ in the $s$ channel.
Also note that the $s$-channel diagram with a NC insertion for the
triple gauge boson vertex depends on the representation chosen in the
gauge sector and therefore satisfies the Ward identity individually.

We are left with the following four diagrams:
\begin{equation*}
\parbox{46mm}{\includegraphics{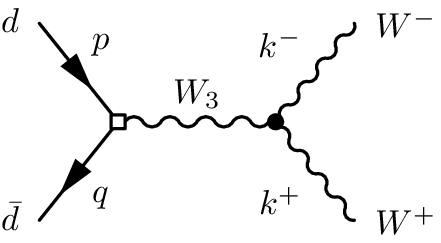}} =
 \ii\mathcal{M}_{u\bar u/d\bar d, s}^{\mu\nu}k^+_\mu\epsilon^-_\nu
\end{equation*}
\begin{equation*}
\parbox{31mm}{\includegraphics{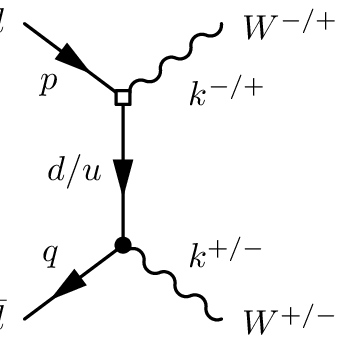}} =
 \ii\mathcal{M}_{u\bar u/d\bar d, t, -/+}^{\mu\nu}k^+_\mu\epsilon^-_\nu
\end{equation*}
\begin{equation*}
\parbox{31mm}{\includegraphics{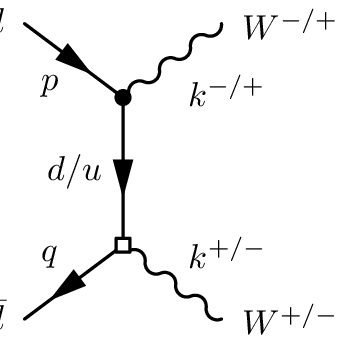}} =
 \ii\mathcal{M}_{u\bar u/d\bar d, t, +/-}^{\mu\nu}k^+_\mu\epsilon^-_\nu
\end{equation*}
\begin{equation*}
\parbox{31mm}{\includegraphics{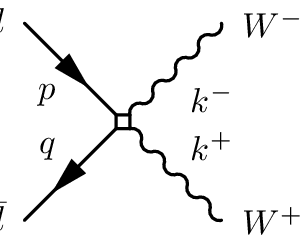}}\quad =
 \ii\mathcal{M}_{u\bar u/d\bar d, 4}^{\mu\nu}k^+_\mu\epsilon^-_\nu\,.
\end{equation*}
Since the choice of incoming vs.~outgoing momenta and particle
vs.~antiparticle is of crucial importance for our argument, we elect
to compute the crossed amplitude with \emph{all} particles, momenta
and quantum numbers \emph{incoming}, as this offers the least
possibility for confusion when applying the Feynman rules.  While this
amplitude does not correspond to a physical process, it is related to
one by crossing symmetry and must therefore satisfy the Ward
identity~\eqref{eq:WI} as well.

Using the Feynman rules of appendix~\ref{app:nmNCSM-Feynman-rules} and
extracting common coupling factors and external fermion wave functions
\begin{equation}
  \mathcal{M}_i^{\mu\nu}k^+_\mu\epsilon^-_\nu =
     \lambda\frac{g^2}{4} \bar{v}(q)\Gamma_i\frac{1-\gamma^5}{2}u(p)
\end{equation}
we find for the individual contributions
\begin{subequations}
\label{eq:WI-diagrams}
\begin{align}
  \Gamma_{u\bar u, s}    &=    (q\theta p) \slashed{\epsilon}_-    \\
  \Gamma_{d\bar d, s}    &=    (p\theta q) \slashed{\epsilon}_-    \\
  \Gamma_{u\bar u, t, -} &=    (k_-\theta p) \slashed{\epsilon}_- 
                             - (\epsilon_-\theta p) \slashed{k}_-  \\
  \Gamma_{d\bar d, t, -} &=    (k_-\theta q) \slashed{\epsilon}_-
                             - (\epsilon_-\theta q) \slashed{k}_- \\
  \Gamma_{u\bar u, t, +} &=  0 \\
  \Gamma_{d\bar d, t, +} &=  0 \\
  \Gamma_{u\bar u, 4}    &=    (k_+\theta p) \slashed{\epsilon}_-
                             - (\epsilon_-\theta p) \slashed{k}_+ \\
  \Gamma_{d\bar d, 4}    &=    (k_+\theta q) \slashed{\epsilon}_-
                             - (\epsilon_-\theta q) \slashed{k}_+ 
\end{align}
\end{subequations}
after using momentum conservation, mass shell conditions,
transversality and the Dirac equations
\begin{subequations}
\label{eq:momenta-etc}
\begin{align}
  p + q + k^+ + k^- &= 0\\
  p^2 = q^2 = {k^+}^2 = {k^-}^2 &= 0 \\
  k^-\epsilon^- &= 0 \\
  \slashed{p} u(p) &= 0 \\
  \bar v(q) \slashed{q} &= 0
\end{align}
\end{subequations}
repeatedly. Using~\eqref{eq:momenta-etc} again, we can verify
from~\eqref{eq:WI-diagrams} the Ward identities in both cases
\begin{subequations}
\begin{align}
    \Gamma_{u\bar u, s}
  + \Gamma_{u\bar u, t, -}
  + \Gamma_{u\bar u, t, +}
  + \Gamma_{u\bar u, 4} &= 0\\
    \Gamma_{d\bar d, s}
  + \Gamma_{d\bar d, t, -}
  + \Gamma_{d\bar d, t, +}
  + \Gamma_{d\bar d, 4} &= 0\,.
\end{align}
\end{subequations}
However we also see that the individual contributions differ, in
particular
\begin{equation}
  \Gamma_{u\bar u, 4} \not=
  \Gamma_{d\bar d, 4}\,,
\end{equation}
even if the relations~\eqref{eq:momenta-etc} are taken into account.

This completes our proof that the different momentum dependence of the
quartic couplings for quarks with isospin up and down, as obtained
from the Lagrangian in~\cite{Melic:2005fm} but missing from the
Feynman rules given in the same publication, is required by gauge
invariance.  It has to cancel unphysical contributions that have a
different momentum dependence, because charge conservation exchanges a
$t$-channel diagram in~$\mathcal{M}^{\mu\nu}_{u\bar u, t, -}$ by a
$u$-channel diagram in~$\mathcal{M}^{\mu\nu}_{d\bar d, t, -}$.
\end{appendix}


\begin{thebibliography}{99}
\bibitem{Seiberg:1999vs}
  N.~Seiberg and E.~Witten,
  %``String theory and noncommutative geometry,''
  JHEP {\bf 9909}, 032 (1999)
  [arXiv:hep-th/9908142].
  %%CITATION = JHEPA,9909,032;%%
\bibitem{Madore:2000en}
  J.~Madore, S.~Schraml, P.~Schupp and J.~Wess,
  %``Gauge theory on noncommutative spaces,''
  Eur.\ Phys.\ J.\  C {\bf 16}, 161 (2000)
  [arXiv:hep-th/0001203];
  %%CITATION = EPHJA,C16,161;%%
%bibitem{Jurco:2001rq}
  B.~Jurco, L.~Moller, S.~Schraml, P.~Schupp and J.~Wess,
  %``Construction of non-Abelian gauge theories on noncommutative spaces,''
  Eur.\ Phys.\ J.\  C {\bf 21}, 383 (2001)
  [arXiv:hep-th/0104153].
  %%CITATION = EPHJA,C21,383;%%
\bibitem{Calmet:2001na}
  X.~Calmet, B.~Jurco, P.~Schupp, J.~Wess and M.~Wohlgenannt,
  %``The standard model on non-commutative space-time,''
  Eur.\ Phys.\ J.\  C {\bf 23}, 363 (2002)
  [arXiv:hep-ph/0111115].
  %%CITATION = EPHJA,C23,363;%%
\bibitem{Melic:2005fm}
  B.~Melic, K.~Passek-Kumericki, J.~Trampetic, P.~Schupp and M.~Wohlgenannt,
  %``The standard model on non-commutative space-time: Electroweak currents  and
  %Higgs sector,''
  Eur.\ Phys.\ J.\  C {\bf 42}, 483 (2005)
  [arXiv:hep-ph/0502249];
  %%CITATION = EPHJA,C42,483;%%
  % B.~Melic, K.~Passek-Kumericki, J.~Trampetic, P.~Schupp and M.~Wohlgenannt,
  %``The standard model on non-commutative space-time: Strong interactions
  %included,''
  Eur.\ Phys.\ J.\  C {\bf 42}, 499 (2005)
  [arXiv:hep-ph/0503064].
  %%CITATION = EPHJA,C42,499;%%
\bibitem{Alboteanu:2006hh}
  A.~Alboteanu, T.~Ohl and R.~R\"uckl,
  %``Probing the noncommutative standard model at hadron colliders,''
  Phys.\ Rev.\  D {\bf 74}, 096004 (2006)
  [arXiv:hep-ph/0608155].
  %%CITATION = PHRVA,D74,096004;%%
\bibitem{Hinchliffe:2002km}
  I.~Hinchliffe, N.~Kersting and Y.~L.~Ma,
  %``Review of the phenomenology of noncommutative geometry,''
  Int.\ J.\ Mod.\ Phys.\  A {\bf 19}, 179 (2004)
  [arXiv:hep-ph/0205040].
  %%CITATION = IMPAE,A19,179;%%
\bibitem{Behr:2002wx}
  W.~Behr, N.~G.~Deshpande, G.~Duplancic, P.~Schupp, J.~Trampetic and J.~Wess,
  %``The Z --> gamma gamma, g g decays in the noncommutative standard model,''
  Eur.\ Phys.\ J.\  C {\bf 29}, 441 (2003)
  [arXiv:hep-ph/0202121].
  %%CITATION = EPHJA,C29,441;%%
\bibitem{Ohl:2004tn}
  T.~Ohl and J.~Reuter,
  %``Testing the noncommutative standard model at a future photon collider,''
  Phys.\ Rev.\  D {\bf 70} (2004) 076007
  [arXiv:hep-ph/0406098].
  %%CITATION = PHRVA,D70,076007;%%
\bibitem{Kilian:2007gr}
  W.~Kilian, T.~Ohl and J.~Reuter,
  %``WHIZARD: Simulating Multi-Particle Processes at LHC and ILC,''
  arXiv:0708.4233 [hep-ph].
  %%CITATION = ARXIV:0708.4233;%%
%%% THIS IS MUCH OLDER ....
%%% \bibitem{Affolder:1999mp}
%%%   A.~A.~Affolder {\it et al.}  [CDF Collaboration],
%%%   %``Measurement of the helicity of $W$ bosons in top quark decays,''
%%%   Phys.\ Rev.\ Lett.\  {\bf 84}, 216 (2000)
%%%   [arXiv:hep-ex/9909042].
%%%   %%CITATION = PRLTA,84,216;%%
\bibitem{Asakawa:1999cu}
  T.~Asakawa and I.~Kishimoto,
  %``Comments on gauge equivalence in noncommutative geometry,''
  JHEP {\bf 9911}, 024 (1999)
  [arXiv:hep-th/9909139].
  %%CITATION = JHEPA,9911,024;%%
\bibitem{Aschieri:2002mc}
  P.~Aschieri, B.~Jurco, P.~Schupp and J.~Wess,
  %``Non-commutative GUTs, standard model and C, P, T,''
  Nucl.\ Phys.\  B {\bf 651}, 45 (2003)
  [arXiv:hep-th/0205214].
  %%CITATION = NUPHA,B651,45;%%
\bibitem{Alboteanu:2007bp}
  A.~Alboteanu, T.~Ohl and R.~R\"uckl,
  %``The Noncommutative Standard Model at $O(\theta^2)$,''
  Phys.\ Rev.\  D {\bf 76} (2007) 105018
  [arXiv:0707.3595 [hep-ph]].
  %%CITATION = PHRVA,D76,105018;%%
\bibitem{SheikhJabbari:2000vi}
  M.~M.~Sheikh-Jabbari,
  %``Discrete symmetries (C,P,T) in noncommutative field theories,''
  Phys.\ Rev.\ Lett.\  {\bf 84}, 5265 (2000)
  [arXiv:hep-th/0001167].
  %%CITATION = PRLTA,84,5265;%%
%%% \bibitem{Douglas:2001ba}
%%%   M.~R.~Douglas and N.~A.~Nekrasov,
%%%   %``Noncommutative field theory,''
%%%   Rev.\ Mod.\ Phys.\  {\bf 73}, 977 (2001)
%%%   [arXiv:hep-th/0106048].
%%%   %%CITATION = RMPHA,73,977;%%
\bibitem{Vermaseren:2000nd}
  J.~A.~M.~Vermaseren,
  %``New features of FORM,''
  [arXiv:math-ph/0010025].
  %%CITATION = MATH-PH/0010025;%%
\bibitem{Moretti:2001zz}
  M.~Moretti, T.~Ohl and J.~Reuter,
  %``O'Mega: An optimizing matrix element generator,''
  [arXiv:hep-ph/0102195].
  %%CITATION = HEP-PH/0102195;%%
\bibitem{Lai:1999wy}
  H.~L.~Lai {\it et al.}  [CTEQ Collaboration],
  %``Global {QCD} analysis of parton structure of the nucleon: CTEQ5 parton
  %distributions,''
  Eur.\ Phys.\ J.\  C {\bf 12}, 375 (2000)
  [arXiv:hep-ph/9903282].
  %%CITATION = EPHJA,C12,375;%%
\bibitem{Conley:2008kn}
  J.~A.~Conley and J.~L.~Hewett,
  %``Effects of the Noncommutative Standard Model on WW scattering,''
  [arXiv:0811.4218 [hep-ph]].
  %%CITATION = ARXIV:0811.4218;%%
\bibitem{Ohl:2008ri}
  T.~Ohl and C.~Speckner,
  %``Production of Almost Fermiophobic Gauge Bosons in the Minimal Higgsless
  %Model at the LHC,''
  Phys.\ Rev.\  D {\bf 78} (2008) 095008
  [arXiv:0809.0023 [hep-ph]].
  %%CITATION = PHRVA,D78,095008;%%
\bibitem{Boos:1999qc}
  E.~Boos and T.~Ohl,
  %``Minimal gauge invariant classes of tree diagrams in gauge theories,''
  Phys.\ Rev.\ Lett.\  {\bf 83} (1999) 480
  [arXiv:hep-ph/9903357];
  %%CITATION = PRLTA,83,480;%%
%bibitem{Ohl:2003jr}
  T.~Ohl and C.~Schwinn,
  %``Forests, groves and Higgs bosons: Gauge invariance classes in
  %spontaneously broken gauge theories,''
  Eur.\ Phys.\ J.\  C {\bf 30}, 567 (2003)
  [arXiv:hep-ph/0305334].
  %%CITATION = EPHJA,C30,567;%%
\end{thebibliography}
\end{document}